# PRELIMINARY RESULTS FROM THE ESO SLICE PROJECT (ESP)

G. Vettolani and E. Zucca
*Istituto di Radioastronomia del CNR, Bologna, Italy*

A. Cappi, R. Merighi, M. Mignoli, G. Stirpe and G. Zamorani
*Osservatorio Astronomico di Bologna, Bologna, Italy*

H. MacGillivray
*Royal Observatory Edinburgh, Edinburgh, United Kingdom*

C. Collins
*Liverpool John Moores University, Liverpool, United Kingdom*

C. Balkowski, J. Alimi, A. Blanchard, V. Cayatte,
P. Felenbok, S. Maurogordato and D. Proust
*DAEC, Observatoire de Paris-Meudon, Meudon, France*

G. Chincarini and L. Guzzo
*Osservatorio Astronomico di Brera, Milano, Italy*

D. Maccagni
*Istituto di Fisica Cosmica e Tecnologie Relative, Milano, Italy*

R. Scaramella
*Osservatorio Astronomico di Roma, Monteporzio Catone, Italy*

M. Ramella
*Osservatorio Astronomico di Trieste, Trieste, Italy*

## ABSTRACT

We present the first results of a galaxy redshift survey, ESO Slice Project (ESP), we are accomplishing as an ESO Key-Project over about 30 square degrees in a region near the South Galactic Pole. The limiting magnitude is $b_J = 19.4$. Observations have been almost completed and $\sim 90\%$ of the data obtained so far has been reduced providing $\sim 3000$ galaxy redshifts. We present some preliminary results concerning the large scale galaxy distribution and their luminosity function.

## 1 Introduction

In 1991 we started a galaxy redshift survey over a strip of $22° \times 1°$ (plus a near-by area of $5° \times 1°$, five degrees west of the strip) in the South Galactic Pole region. The right ascension limits are $22^h30^m$ and $01^h20^m$, at a mean declination of $-40°$ (1950). We have filled this area with a regular grid of circular fields with a diameter of 32 *arcmin*.

We have selected the target objects from the Edinburgh–Duhram Southern Sky Galaxy Catalogue[1]. The limiting magnitude of the survey is $b_J \leq 19.4$.

At $z \simeq 0.1$, which corresponds to the peak of the selection function of the survey, the linear dimensions of the full strip are of the order of $110 \times 5 \ h^{-1} \ Mpc$.

## 2  Observations and Data Reduction

The observations have been obtained with the multifiber spectrograph OPTOPUS at the Cassegrain focus of the ESO 3.6$m$ telescope at La Silla. OPTOPUS has 50 fibers (2.4 *arcseconds* diameter) which are manually plugged into holes drilled in aluminum plates.

The spectra cover the wavelength range from 3730 Å to 6050 Å, with an effective resolution of about 3.5 Å. Spectra are flat–fielded, extracted and wavelength calibrated. Then the relative transmission of each fiber is computed by normalizing each spectrum through the flux of the OI $\lambda 5577$ and Na $\lambda 5891$ sky lines.

After subtraction of the average sky from the galaxy plus sky spectra (the accuracy of the sky subtraction is in the range 2–5%), the redshifts are measured by cross–correlating the spectra with a set of template stars observed by us with the same instrument. The redshifts from emission lines (when present) are also measured, as well as the equivalent width of the most common emission lines. The median error in velocity for the total sample is $\sim 50 \ km/s$.

About 70% of the fields observed up to now have a redshift completeness greater than 80%. The next observing runs will be used to higher the completeness of the fields.

## 3  Large Scale Properties

The histogram in Figure 1 shows the distribution in comoving distance ($q_\circ = 0.5$) of the 2856 galaxies measured so far.

The outstanding peak at $D \simeq 300 \ h^{-1} \ Mpc$ is not due to a single galaxy cluster but to a structure which extends over almost all the fields. The distance of this peak is almost coincident with the maximum of our selection function. Other peaks seen in Figure 1 at smaller and larger distance are due to less prominent structures, but with a similar contrast with respect to a uniform distribution.

A large fraction of galaxies ($\sim 50\%$) shows the presence of one or more emission lines (OII $\lambda 3727$, $H\beta$, OIII $\lambda 4959$ and $\lambda 5007$). These objects can be either spiral galaxies, where lines originate mostly from HII regions in the disks, or early–type galaxies undergoing a significant burst of star formation.

The large scale distribution of galaxies with emission lines is different from that of galaxies without emission lines: the observed peaks in the galaxy distribution are much less apparent in galaxies with emission lines. This suggests that either spiral galaxies are less frequent in the densest regions, thus confirming a large scale validity of the well known morphology–density relation, or starbursts phenomena occur preferentially in low density environments or both. After completion of the survey we will be able to explore further this point on the basis of quantitative estimates of densities and a clearer definition of the structures beyond pure visual impression.

Figure 1: Galaxy distribution in comoving distance ($q_0 = 0.5$). The solid line shows the expectation resulting from a uniform distribution of the galaxies in the sample. Vertical lines show the position of the BEKS peaks (see text).

The vertical lines in Figure 1 show the location of the regularly spaced density enhancements found in deep pencil beam survey (BEKS)in the South Galactic Pole region[2,3]. The two main peaks in our redshift distribution (at comoving distances of $\sim 170$ and $\sim 300\ h^{-1}\ Mpc$) are in reasonably good agreement with the BEKS peaks and may well be part of the same structures (walls) orthogonal to the line of sight.

Under this hypothesis, since the BEKS[3] survey region is located $\sim 10°$ north of the eastern corner of the present survey, the structure at $z \simeq 0.1$ would have minimum linear dimensions of the order of $110 \times 50\ h^{-1}\ Mpc$, comparable with the Great Wall[4].

Although really apparent only in the NGP–SGP direction the striking "periodicity" seen in this 1D deep survey has prompted several speculations and some controversy as well. The latter mainly based on the criticism on the estimates of levels of statistical significance of the signal found by BEKS to be at $\lambda = 128 h^{-1}$ Mpc, due to correlated structures and on the presence of aliasing of power because of the lower dimensionality of the BEKS surveys[5].

Even though we cover much less depth than the BEKS survey, it is worthwhile to use the present data to assess the possible presence of peaks in the power spectrum, since in our survey we gain one dimension and this is likely to affect the level of power aliasing. Deferring a more accurate analysis after the completion of the survey, we present in Figure 2 the power spectrum (periodogram) of the radial distribution of the raw data, computed from the $N$ measured redshifts as

$$P(k) \;=\; \frac{1}{N}\left\{\left[\sum_i^N sin(kr_i)\right]^2 + \left[\sum_i^N cos(kr_i)\right]^2\right\}$$

Figure 2: Periodogram of the radial distribution of the redshift data.

In Figure 2 are also shown the probabilities that given levels of $P$ for an uniform distribution[6]. The latter are only indicative, since in the raw data are present effects due to the selection function, redshift space distortions and we know that galaxy are correlated; however, the selection function mainly affects values of $k$ smaller than that of interest ($\lambda_{FWHM} \simeq 700h^{-1}$ Mpc), and redshift disortions affect higher values (see peaks in $P$ at $\lambda \sim 30 - 40h^{-1}$ Mpc).

From Figure 2 is evident that the most prominent peak agrees well with the BEKS value, although is not very narrow. Even at this preliminary stage we conclude that this kind of signal is present in the volume spanned by our data, and its presence has not been weakened during our data gathering: it was already present when we had only $\sim 600$ galaxy redshifts, coming form a sparse subset of our target fields. At the present moment, however, to generalize this result to other sky areas or to infer from it ensemble properties is unwarranted. This notwithstanding, we feel the BEKS original result probably may point out to the possible importance of phases of cosmic fluctuations, which are typically overlooked and averaged out in standard analyses like the two point correlation function.

## 4 The Luminosity Function

The well controlled selection of our sample and the large number of redshifts already obtained allow us to estimate for the first time the shape parameters of the galaxy luminosity function at magnitudes as faint as $b_J = 19.4$.

Since our database was selected in the blue–green, K–corrections are needed to compute the luminosity function even for the moderate redshifts sampled by our galaxies ($z \leq 0.2$). The functional forms of the K–correction as a function of redshift[7] obviously requires the knowledge of the morphological type of each galaxy. To overcome this problem, we have adopted the following statistical approach. First, we have assumed that the percentages of late– and early-type galaxies in our sample are approximately the same as those observed in brighter and nearer samples[7], then we have applied the K–correction appropriate for a late–type galaxy to all galaxies showing emission lines. Finally, we have randomly assigned a morphological type to the remaining galaxies in such a way as to obtain the assumed ratio of late and early types in the total sample.

Following the STY method[8], we have then derived the parameters of the Schechter functional form of the luminosity function, through a maximum likelihood technique. The best fit parameters are $\alpha = -1.13$ and $M^*_{b_J} = -19.70$.

In order to estimate the maximum amount of uncertainty induced by our "statistical" K–correction scheme, we have also computed the two parameters of the luminosity function with the extreme assumptions that all galaxies are either ellipticals or spirals. The best fit parameters obtained in these way are $\alpha = -1.15$, $M^*_{b_J} = -19.87$ and $\alpha = -1.05$, $M^*_{b_J} = -19.56$, respectively.

Figure 3 shows the luminosity function (with arbitrary normalization) obtained for our sample: the solid line corresponds to the Schechter functional form obtained through the STY method, while the solid circles have been determined through a modified version of the non–parametric C–Method[9]. The 68% and 95% confidence ellipses for $\alpha$ and $M^*_{b_J}$ are shown in the inset, where the cross represents the best estimate obtained[10] from a survey of galaxies with $b_J \leq 17.15$.

The excellent agreement shows that, up to a depth of $\simeq 600\ h^{-1}\ Mpc$, the overall shape of the luminosity function is well described by a Schechter function with a faint–end slope $\alpha \sim -1.1$ up to $M = -16$.

## 5 Acknowledgements

This work, based on data collected at the European Southern Observatory, has been partially supported through NATO Grant CRG 920150 and EEC Contract ERB–CHRX–CT92–0033.

Figure 3: Galaxy luminosity function: solid circles have been computed using a modified version of the C–method and the normalization is arbitrary.

# 6 References

1. Heydon–Dumbleton, N.H., Collins, C.A., MacGillivray, H.T., 1988, in *Large–Scale Structures in the Universe*, ed. W. Seitter, H.W. Duerbeck and M. Tacke (Springer–Verlag), p. 71.
2. Broadhurst, T.J., Ellis, R.S., Shanks, T., 1988, MNRAS 235, 827
3. Broadhurst, T.J., Ellis, R.S., Koo, D.C., Szalay, A.S., 1990, Nature 343, 726
4. Geller, M.J., Huchra, J.P., 1989, Science 246, 897
5. Kaiser, N, Peacock, J.A., 1991, Ap.J. 379, 482
6. Duari, D., Das Gupta, P., Narlikar, J., 1992, Ap.J. 384, 35
7. Shanks, T., Stevenson, P.R.F., Fong, R., MacGillivray, H.T., 1984, MNRAS 206, 767
8. Sandage, A., Tamman, G., Yahil, A., 1979, Ap.J. 232, 252
9. Lynden–Bell, D., 1971, MNRAS 155, 95
10. Loveday, J., Peterson, B.A., Efstathiou, G., Maddox, S.J., 1992, Ap.J. 390, 338